\documentclass[useAMS,usenatbib,usegraphicx]{mn2e}

\title[AGB envelopes as probes of binary companions]{The shapes of AGB
 envelopes as probes \\of binary companions}
\author[P. J. Huggins et al.]{P. J. Huggins$^{1}$\thanks{E-mail:
patrick.huggins@nyu.edu (PJH); mauron@graal.univ-montp2.fr (NM);
 eaw300@nyu.edu (EAW)}, N. Mauron$^{2}$, and E. A. Wirth$^{1}$
\footnotemark[1]\thanks{Based in part on data from the
 NASA/ESA/CADC HST archives, and the ESO VLT archive. }\\
\
$^{1}$Physics Department, New York University, 4 Washington Place, New York
NY 10003, USA \\
$^{2}$GRAAL, CNRS, and Universit{\'e} Montpellier II,
 Place Bataillon, 34095 Montpellier, France \\
}
\begin{document}

\date{Accepted ... Received ....; in original form ...}

\pagerange{\pageref{firstpage}--\pageref{lastpage}} \pubyear{2008}

\maketitle

\label{firstpage}

\begin{abstract}
We describe how the large scale geometry of the circumstellar
envelopes of asymptotic giant branch stars can be used to probe the
presence of unseen stellar companions.  A nearby companion modifies
the mass loss by gravitationally focusing the wind towards the orbital
plane, and thereby determines the shape of the envelope at large
distances from the star.  Using available simulations, we develop a
prescription for the observed shapes of envelopes in terms of the
binary parameters, envelope orientation, and type of observation. The
prescription provides a tool for the analysis of envelope images at
optical, infrared, and millimetre wavelengths, which can be used to
constrain the presence of companions in well observed cases. We
illustrate this approach by examining the possible role of binary
companions in triggering the onset of axi-symmetry in planetary nebula
formation.  If interaction with the primary leads to
axi-symmetry, the spherical halos widely seen around newly formed
nebulae set limits on the companion mass. Only low mass objects may
orbit close to the primary without observable shaping effects: they
remain invisible until the interaction causes a sudden change in
the mass loss geometry.
\end{abstract}


\section{Introduction}

One of the striking aspects of the evolution of stars from the
asymptotic giant branch (AGB) to the planetary nebula (PN) phase is
the sudden change in morphology of the circumstellar gas. On the AGB,
the stellar mass loss is thought to be roughly spherically symmetric,
but this rapidly evolves into prominent axi-symmetry on evolution into
the proto-PN phase.  The axial symmetry may consist of a smooth
gradient in mass loss from pole to equator or a distinct equatorial
torus, and is often accompanied by high velocity bi-polar or
multi-polar jets, aligned approximately along the symmetry axis.

Most of the theoretical ideas that have been proposed to explain the
axi-symmetry rely on the presence of a binary companion. There is,
however, no consensus on a specific scenario, and exactly what happens
is controversial. Proposals include tidal effects; engulfment of the
companion with expulsion of a common envelope; and the effects of
jets, from accretion disks around the companion or around the core of
the primary \citep[e.g.,][]{mor87,sok94,rey99,sok00,nor06}.

If the axial symmetry of PNe is due to the effects of a binary
companion, the relatively common occurrence of this phenomenon
\citep[e.g.,][]{sah07} strongly constrains the possible star systems
that become PNe. However, the observational situation with respect to
binarity is not clear.  Relatively few AGB stars are known to have
binary companions \citep[Table 9.1]{jor03}. Unless the companion is
bright or hot, it is difficult to detect given the high luminosity and
variability of the primary and its thick circumstellar envelope. The
central stars of PNe are somewhat more accessible targets, and
extensive searches have been made for companions using photometry,
imaging, and radial velocity measurements (e.g., \citealt{dem06,
mis09}). However, only a few dozen cases with companions have been
found. These are typically short period systems which imply direct
interaction in the past, but the general issue of PN formation remains
an open question, especially in view of the difficulty in finding
companions with intermediate or long periods.

In this paper we discuss a different approach for probing the presence
of a binary companion that might affect the AGB--PN transition. It is
based on the influence of the companion in shaping the mass loss of
the AGB star before the transition takes place. The shaping effect is
frozen into the expanding circumstellar envelope, and is reflected in
its large scale geometry. 

This approach is possible because changes in the separation of the
system are reasonably well prescribed. In cases where the companion
comes into contact with or is engulfed by the AGB star during the
transition, it must initially have been relatively close, within
several AU. If the companion induces the transition from somewhat
farther out, the orbital separation is expected to undergo a slow
increase in response to the mass loss of the system. In all cases
where the companion significantly affects the transition, it is likely
to be well within the circumstellar envelope during the AGB phase, and
can be expected to affect the large scale geometry of the envelope
over a long time period.  The geometry of the extended envelope can be
studied with a variety of techniques on the AGB. It can also be
studied in the halos of proto-PNe and PNe, in material that was
ejected earlier by the AGB star and lies outside the developing
axi-symmetric nebula.

The plan of the paper is as follows. In the next section we discuss
the effects of detached binary companions on the large scale geometry
of AGB envelopes.  We then develop a prescription for the observed
shapes of the envelopes in terms of the binary parameters and the
orientation.  We illustrate how these results can be used to probe the
presence of companions by measuring the shape in well-studied cases,
and we discuss some implications for the formation of PNe.

\section{Shaping effects of an AGB companion}

There are no systematic observations of the extended envelopes of AGB
stars with companions, although there are a few cases known in which
strong interactions already appear to be taking place. For example,
$\pi^1$ Gru shows an equatorial torus and jets, similar to the
configuration seen in proto-PNe \citep[e.g.,][]{chi06,sac08,hug07}.  We
therefore rely on numerical simulations to understand the effects of a
companion on the large scale structure (\citealt{the93};
\citealt{mas99} [hereafter MM99]; \citealt{gaw02}). The available
simulations are not very extensive, but the general nature of the
effects caused by the companion is clear. In addition to local
interactions which may include tidal spin-up of the primary and
accretion of envelope material by the secondary, there are two main
effects on the extended circumstellar envelope.

First, for a wide range of conditions, a spiral pattern is imprinted
on the envelope, mainly as a result of the reflex motion of the
primary. An example of this has been detected in the envelope of the
evolved carbon star AFGL~3068 \citep{mau06,mor06}.  The spiral formed
by this mechanism is not easy to detect. The radial wavelength of the
pattern ($\lambda$) is given by $\lambda = VP$, where $V$ is the
expansion velocity of the envelope, and $P$ is the period of the
binary. Even for relatively nearby AGB stars the pattern is resolvable
only for long periods. In the case of AFGL~3068 the period is
$\sim$~800~yr and the corresponding separation $\sim$ 100~AU, which is
probably too large for the companion to play an important role in the
AGB-PN transition.

The second effect of a companion on the extended envelope is to modify
the mass loss geometry by gravitationally focusing matter towards the
orbital plane.  The latitude-density profile is eventually frozen into
the wind flow at the terminal velocity in each direction, and this
results in a flattened large scale structure.  The evolutionary time
scale on the AGB is much longer than the binary orbital periods of
interest here, so the spiral pattern, if present, forms a fine
structure on the extended envelope. Averaged over many spiral
features, the large scale structure provides a probe of the presence
of the companion, with a characteristic size scale much larger than
the binary separation or the spiral pattern. It is this large scale
structure that we quantify in the following sections.

\section{The envelope shape}

\subsection{Characterising the shape}

We develop an approximate prescription for the large scale binary
shaping of AGB envelopes based on the binary simulations of MM99,
which are the most comprehensive. Details of the numerical techniques,
physical assumptions, and limitations of the simulations are discussed
at length by \cite{mas98}. The models we use are listed in Table~A1.  They
employ a single primary mass of 1.5~M$_\odot$ with mass loss rates
$\sim$ 1--$2 \times 10^{-5}$~M$_\odot$~yr$^{-1}$ appropriate for AGB
stars. The parameter space covers 0.25--2.0~M$_\odot$ for the
secondary mass ($M_{\rm s}$), 3.6--50.4~AU for the binary separation ($d$),
and 5--17~km~s$^{-1}$ for the wind velocity at the location of the
secondary ($V_{\rm s}$).  The wind from the star is assumed to be initially
spherically symmetric. A close-in secondary may spin-up the primary or
cause other effects which may enhance equatorial mass loss. In this
sense the simulations provide a lower limit to the actual degree of
axi-symmetry.

The overall shapes of the envelopes affected by gravitational focusing
vary from model to model, depending on the parameters of the
binary. For high mass, close-in companions, the wind is highly concentrated
towards the orbital plane, and for lower mass, more distant companions
the flow is more spherically symmetric. At large distances from the
star system where the wind is in steady state at the terminal velocity, the
latitude variation of the density is frozen into the flow.

We characterise the latitude variation of density in the extended
envelope using an exponential profile.  The density is then given by
the equation:
\begin{equation}
n(r,\theta) = \frac{n_{\rm a} r^2_{\rm a}}{r^2} \exp(-\theta/\theta_{\rm o}),
\end{equation}
where the $r$ is the distance from the centre, $n_{\rm a}$ is the equatorial
density at reference radius $r_{\rm a}$, $\theta$ is the latitude (measured
from the orbital plane in degrees), and $\theta_{\rm o}$ is the scale
height.  The exponential profile is adopted as a simple approximation
to the actual latitude variation of the five simulations reported by
MM99 in their Fig.~23 and Table~3. These profiles differ in detail
from model to model; the procedure used below matches them to an
exponential at the orbital plane and at intermediate latitudes.  For
the most flattened envelope (model~1), the approximation is not an
accurate fit of the detailed profile near the plane: the mean (absolute)
deviation (averaged over latitude $0\degr$--$85\degr$) is 0.19~dex. In
this case however, the total variation in density from pole to equator
is extremely large, more than a factor of 300 (2.6~dex), so the
approximation is still a useful descriptor of the overall shape. For
the less flattened envelopes, which are the main focus of later
discussion, the procedure provides quite good approximations to the
detailed profiles.  For example, for model 3, the exponential differs
from the exact profile by a mean (absolute) deviation of 0.03~dex. 

To determine the value of $\theta_{\rm o}$ for each simulation in Table~A1
we fit equation~(1) to the ratio of the density at the equator and
intermediate latitudes 50\degr\ and 80\degr, and adopt the geometric
mean. The dispersion of the individual estimates about the adopted
values is $10\%$ (rms).

A shape parameter equivalent to $\theta_{\rm o}$ is the density
contrast $(n_{\rm eq}-n_{\rm po})/n_{\rm po}$, where $n_{\rm eq}$ and $n_{\rm po}$ are the
densities at the equator and poles, at the same radial distance. It is
convenient to write the contrast as $K_n-1$ where $K_n$ is the
equator/pole density ratio, i.e.,
\begin{equation}
K_n - 1 = (n_{\rm eq}-n_{\rm po})/n_{\rm po}.
\end{equation} 
This is zero for a spherical envelope, and increases with the degree
of flattening.  For the exponential model, it is seen from
equation~(1) that $K_n$ and $\theta_{\rm o}$ are related by the expression
$K_n=\exp(90/\theta_{\rm o})$.  These simple characterisations are used to
represent the overall shape of each envelope.

\subsection{Relation to binary parameters }

\begin{figure}
\includegraphics[width=80mm]{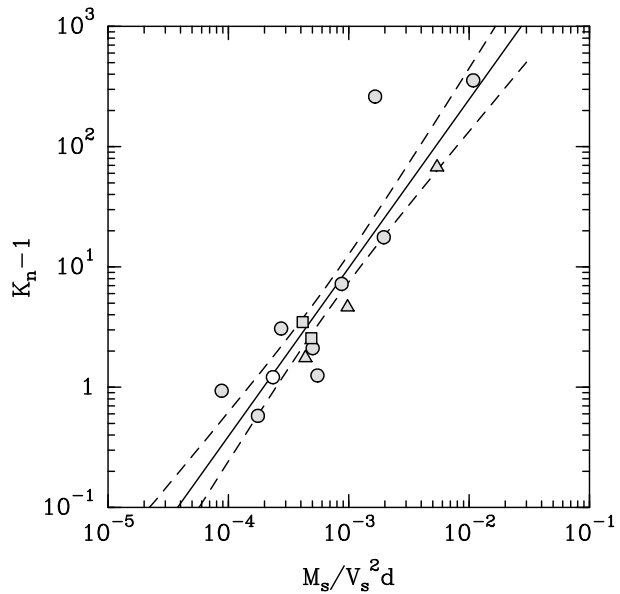}
\caption{Envelope shape parameter $K_n-1$ vs.\ binary parameters
 $M_{\rm s}/V_{\rm s}^2d$, in units of $\rmn{M_\odot/(km~s^{-1})^2\,AU}$. The
 symbols show model data from Table A1 and indicate the value of the
 parameter $1+M_{\rm p}/M_{\rm s}$: open circles (1.75), filled circles (2.5),
 triangles (4.0), and squares (7.0). The solid line is the least
 squares fit to the data, and the dashed lines are one-sigma
 confidence limits of the fit.  }
\label{fig1}
\end{figure}

\begin{figure}
\includegraphics[height=80mm]{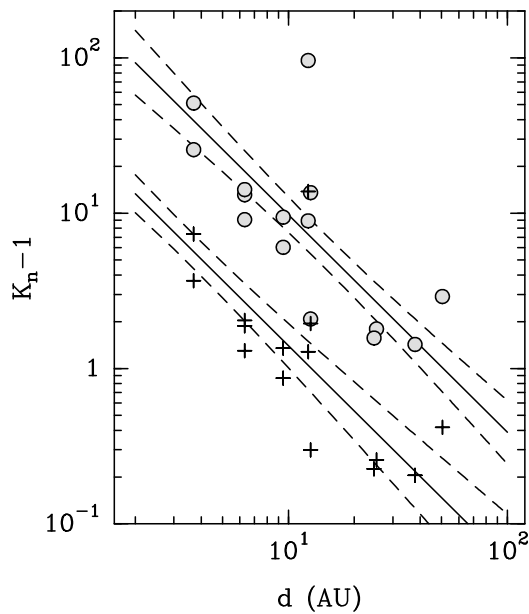}
\caption{Envelope shape parameter $K_n-1$ vs.\ separation $d$,
 projected from Fig.~1 for $V_{\rm s}= 10$~km~s$^{-1}$. The upper lines
 and circles are for $M_{\rm s}=1$~M$_{\sun}$; the lower lines and
 crosses are for $M_{\rm s}= 0.25$~M$_{\sun}$.
}
\label{fig2}
\end{figure}

\begin{figure}
\includegraphics[height=80mm]{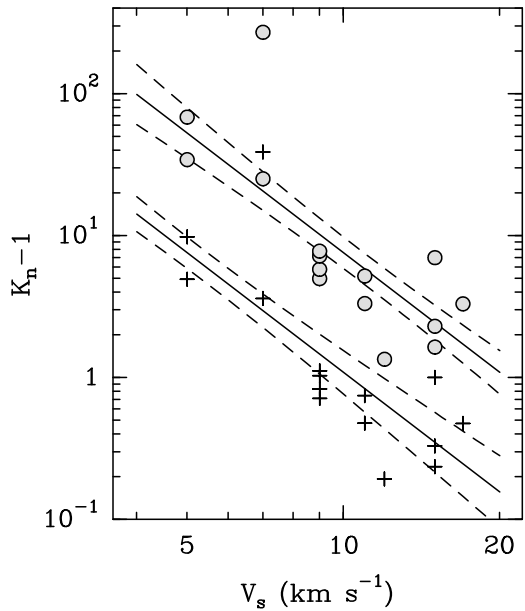}
\caption{Envelope shape parameter $K_n-1$ vs.\ wind velocity $V_{\rm s}$,
 projected from Fig.~1 for separation $d= 12$~AU.  The upper lines
 and circles are for $M_{\rm s}=1$~M$_{\sun}$; the lower lines and
 crosses are for $M_{\rm s}= 0.25$~M$_{\sun}$.
}
\label{fig3}
\end{figure}

The next step in our prescription is to determine an empirical
relation between the binary parameters and the envelope shape, using
the density contrast $K_n-1$ introduced above.  We follow in part the
earlier discussion by MM99.

In order to guide the analysis, we assume that the gravitational
focusing depends only on the following parameters: the masses of the
primary and secondary ($M_{\rm p}$ and $M_{\rm s}$), the separation ($d$), and the
velocity of the wind at the orbit of the secondary ($V_{\rm s}$). We
ignore the possible influence of other quantities such as the
dimensions of the stars, the form of the wind acceleration curve,
etc., which we assume to be of secondary importance. It then follows
from dimensional analysis that the density contrast $K_n-1$ depends on
a function $\Phi$ of at most two dimensionless combinations of the
independent variables and the gravitational constant $G$.

In certain limiting cases, physical arguments provide further
information on the form of $\Phi$. For example, in the limit that $V_{\rm o}
\ll V_{\rm s}$, where $V_{\rm o} $ is the orbital velocity ($V_{\rm o}^2 =
G(M_{\rm p}+M_{\rm s})/d$), the trajectories of the wind particles in the
ballistic limit scale with only one dimensionless parameter,
$GM_{\rm s}/V_{\rm s}^2\,d$. Similarly, in the the limit of large $M_{\rm p}$ where the
orbital velocity is large, the presence of the secondary acts
effectively as a thin ring, and the contrast again depends on the same
dimensionless parameter.  In the intermediate regime of interest here,
the situation is less amenable to simple analysis, but the above
considerations motivate a parameterisation of the form:
\begin{equation}
K_n - 1 = \Phi(GM_{\rm s}/V_{\rm s}^2\,d,\ (M_{\rm p}+M_{\rm s})/M_{\rm s}).
\end{equation}

From the numerical results of the simulations, it is found that $K_n -
1$ shows a strong dependence on the first parameter in equation
(3). The data are shown in Fig.~1. The variation is well approximated
by a power law, and a least squares fit gives:
\begin{equation}
\log (K_n - 1) = 5.19\,(\pm0.66) + 1.40\,(\pm0.21)\,\log (M_{\rm s}/ V_{\rm s}^2
d).
\end{equation}
This equation is compared to the data in Fig.~1. The solid line shows
the fit, and the dashed lines show the one sigma (standard error)
confidence limits. One data point (model 12, Table A1) appears as an
outlier and might be affected by resonance or the residual effects of
a spiral shell.

The numerical results of the simulations show no aditional variation
of $K_n - 1$ with the second parameter of equation (3). This is
illustrated in Fig.~1 using different symbols to denote different
values of $1+M_{\rm p}/M_{\rm s}$, which range from 1.75 to 7.0 
(for $M_{\rm s} = 0.25$--2.0~M$_\odot$ and $M_{\rm
p}=1.5$~M$_\odot$). It can be seen that all points follow essentially
the same curve. We can set an approximate upper limit on the influence
of this parameter by including it in the argument to the power law fit
as $GM_{\rm s}/V_{\rm s}^2\,d \,\times\, (1+M_{\rm p}/M_{\rm
s})^\alpha$. We find that $\alpha = -0.29\pm 0.48$ (or $-0.10\pm 0.33$
if we omit the outlier) indicating little or no dependence on the
second parameter, and therefore little or no dependence on the primary
mass.  Thus, even though the simulations were made for a single value
of $M_{\rm p}$ (1.5~M$_\odot$), we can expect roughly similar results
for other values of $M_{\rm p}$ in the relatively small mass range
($\sim 0.6$ to a few M$_\odot$) relevant to stars in the later stages
of the AGB.

We conclude that equation (4) gives the main parametric dependence of
the envelope shape on the binary parameters in the regime of interest,
and we adopt this relation for the remainder of the paper. The
dispersion of the model points, characterised by the median deviation,
is 0.19 dex. This translates into a dispersion of the observed column
density contrast ($C_{\rm N}-1$ see below) of $\sim 25\%$.  Factors which
may contribute to the dispersion are discussed in appendix A.  To show
the dependence of $K_n - 1$ on individual parameters, Fig.~2 plots
equation (4) as a function of $d$, for fixed values of $M_{\rm s}$ and
$V_{\rm s}$, and the model points scaled to these values using the
equation. Figure~3 shows a similar plot as a function of $V_{\rm s}$, for
fixed $M_{\rm s}$ and $d$.  Although the relations are not tightly
constrained, it can be seen that they are sufficiently well determined
to allow a preliminary investigation of the shaping.

The above results are in accord with the discussion by MM99. They
found that the quantity $R^1_{acc}/d$ is useful for distinguishing
different morphological types of envelopes, where $R^1_{acc}$ is an
effective accretion radius ($=2GM_{\rm s}/V^2_{\rm s}$) using the velocity of the
wind at the secondary. Although $R^1_{acc}/d$ has a strict physical
interpretation in terms of accretion only in the limit $V_{\rm o}\ll V_{\rm s}$,
the parameterisation of $M_{\rm s}$, $V_{\rm s}$, and $d$ is the same as in
equation (4) above. MM99 also noted that an analytic formula for the
ratio of the accretion rate to the mass loss rate, called
$\alpha_{\mathrm{focus}}$ by \cite{han95} and used by them to
characterise the degree of focusing, is, in fact, only a fair
indicator of the morphological type. $\alpha_{\mathrm{focus}}$ may be
written as $(GM_{\rm s}/d)^2/V_{\rm s}V_{\rm r}^3$, where $V_{\rm r}$ is the relative velocity
of the wind at the location of the secondary ($V_{\rm r}^2 =
V_{\rm o}^2+V_{\rm s}^2)$. In the limit $V_{\rm o}\ll V_{\rm s}$, $\alpha_{\mathrm{focus}}$
reduces to a parameterisation equivalent to equation (4), but in the
general case it differs.  We find that a power law fit of
$\alpha_{\mathrm{focus}}$ to $K_n - 1$ is a poorer predictor of the
density contrast than equation (4), with a dispersion $\sim 0.12$~dex
larger. Equation (4) is therefore the preferred relation.  More
extensive simulations over a larger range of parameter space are
needed to investigate this further.

\subsection{Observed shapes}

The final step in making the complete connection between the binary
properties and observations, is to relate the intrinsic shapes of the
envelopes (given above by $K_n-1$) to the observed shapes. This
involves a dependence on the orientation of the envelope, which we
take to be the inclination angle ($i$) of the symmetry axis to the
line of sight, and on the kind of observation. For example, it may
depend on the distribution of column density, emission measure, or
some other quantity, projected onto the plane of the sky.

\begin{figure}
\includegraphics[width=80mm]{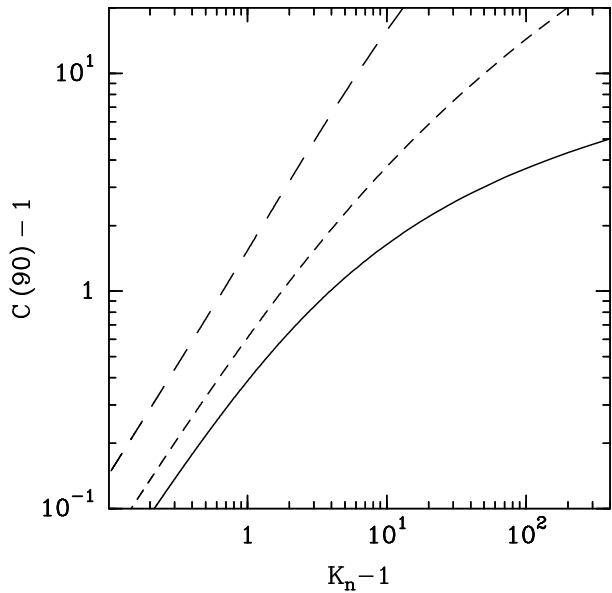}
\caption{Relation between shape parameter $K_n-1$ and intensity contrast
on the major and minor axes for envelopes seen edge-on ($i=90\degr$).
Solid line: $C_{\rm N}$ (intensity $\propto$ column density). Long dashed
line: $C_{\rm E}$ (intensity $\propto$ emission measure). Short dashed line:
$C_{\rm S}$ (scattered central illumination).
}
\label{fig4}
\end{figure}

\subsubsection{Column density}
We first consider the case where the observed intensity depends on
column density. This is relevant to imaging the envelope in optically
thin, dust scattered ambient Galactic light, and other techniques such
as infrared imaging or millimetre molecular line observations which
can be used to determine the column density. We characterise the
observed shape of the envelope by the ratio between the column
density on the projected major and minor axes at the same distance
from the centre, i.e., $C_{\rm N} = N_\rmn{maj}/N_\rmn{min}$.  It is then
straightforward to show that for an edge-on ($i=90\degr$) envelope:
\begin{equation}
C_{\rm N}(90) = \ln K_n \,\, K_n / (K_n - 1).
\end{equation}
This relation is shown as the solid line in Fig.~4.  Highly
non-spherical envelopes with large values of $K_n$ appear flattened,
with $C_{\rm N}(90) \sim \ln K_n$; and nearly spherical envelopes with
values of $K_n$ close to 1, appear nearly circular with $C_{\rm N}(90) \sim
1$.

\begin{figure}
\includegraphics[width=80mm]{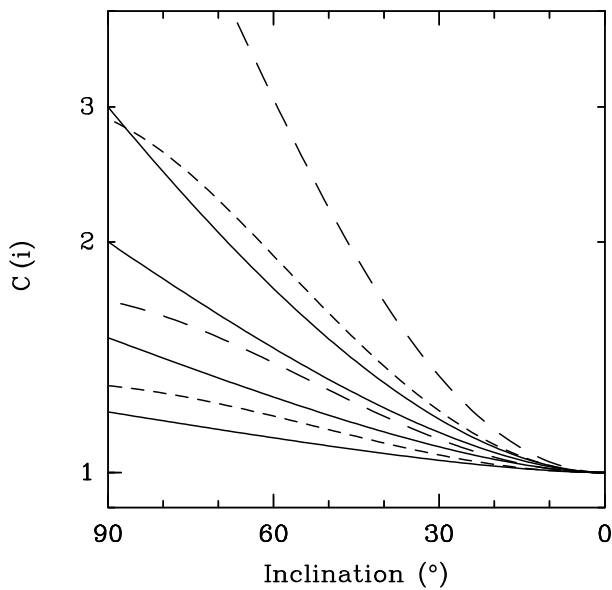}
\caption{Intensity ratio of envelopes as a function of
 inclination angle. Solid lines: $C_{\rm N}$ for values
 of $K_n$ $=$ 1.5, 2.0, 4.9, and 16.8 (bottom to top). Long dashed lines:
 $C_{\rm E}$. Short dashed lines: $C_{\rm S}$. For $C_{\rm E}$ and $C_{\rm S}$,
 $K_n$ $=$ 1.5 and 4.9.
}
\label{fig5}
\end{figure}

For other inclination angles, $C_{\rm N}(i)$ has been evaluated numerically.
The results for several values of $K_n$ are shown by the solid lines
in Fig.~5. In each case $C_{\rm N}(i)$, given by equation~(5) for
$i=90\degr$, decreases with decreasing inclination and reaches 1
(i.e., the envelope appears circular) when the system is seen face-on.

We now have a prescription for the observed shape of an envelope
formed by gravitational focusing in a binary system. Given the binary
parameters $M_s$, $V_{\rm s}$, and $d$, equation (4) determines the
intrinsic structure of the envelope, equation (5) gives the shape seen
edge-on, and Fig.~5 shows how this appears at other inclinations.

For later reference, Fig.~6 shows the variation of $C_{\rm N}(90)$ (the
edge-on case) in the secondary mass--separation plane for a
representative value of $V_{\rm s}$ (10~km~s$^{-1}$).  Lighter lines are used
to extend the contours beyond the actual range of the models
(Table~A1), assuming that equation (4) remains valid. The velocity of
10~km~s$^{-1}$ is the typical terminal velocity of an AGB wind. For
small separations, the velocity at the secondary will be less than
this (depending on the wind acceleration curve) and the values of
$C_{\rm N}$ will be correspondingly larger; the results of specific cases
can be calculated using the equations given.

\begin{figure}
\includegraphics[width=80mm]{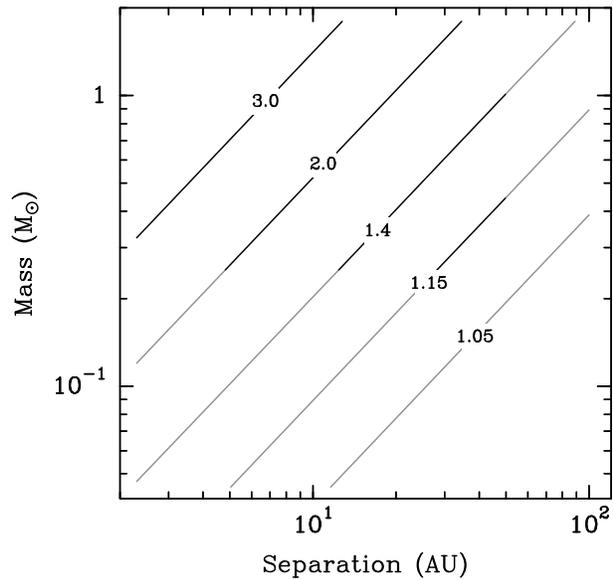}
\caption{Intensity ratio $C_{\rm N} (90)$ (intensity $\propto$ column
density) of envelopes seen edge-on, as a function of the secondary
mass and separation for $V_{\rm s}= 10$~km~s$^{-1}$. The light contours
represent regions beyond the parameter range of the simulations.
}
\label{fig6}
\end{figure}

\subsubsection{Emission measure}
If the image intensity depends on some quantity other than the column
density, the observed shape will in general be different. A dependence
on emission measure arises when the AGB halo of a newly formed PN is
photo-ionised, and is imaged in the continuum or an emission line
whose intensity depends on $\int \!n^2 dl$ along the line of
sight. For this case it may be shown that $C_{\rm E}$, the intensity ratio
on the projected major and minor axes, is related to the shape
parameter $K_n$ by:
\begin{equation}
C_{\rm E}(90) = \ln K_n^2 \, \frac{K_n^2\,(1+4(\ln K_n/\pi)^2)}{K_n^2 -(1+8(\ln
K_n/\pi)^2)},
\end{equation}
for an envelope seen edge-on ($i=90\degr$).  This relation is shown as
the (long) dashed line in Fig.~4. It can be seen that imaging which
depends upon emission measure produces a higher contrast than column
density, because of the increased sensitivity to the density. The
variation of $C_{\rm E}$ with inclination angle has been evaluated
numerically and is found to be similar in form to the curves for
$C_{\rm N}$.  Two examples are shown in Fig.~5.

The shapes of non-spherical AGB halos around PNe are expected to vary
over time. When the envelopes become photo-ionised, gradients in the
gas pressure will produce flows that tend to smooth out any density
gradients \citep{mel95}.  The smoothing time scale at radius $r$ is
$t \sim r/v_{\rm s}$ where $v_{\rm s}$ is the sound speed in the gas. For
$v_{\rm s}=10$~km~s$^{-1}$, $t$ ranges from 3,000 to 30,000~yr for radii of
$10^{17}$ to $10^{18}$~cm, which are the size scales typically
accessible to observations. Thus contrast in precursor AGB envelopes
will still be present in the AGB halos of newly formed PNe, but will
eventually disappear in more evolved nebulae.

\subsubsection{Central illumination}
A variety of other cases can arise when the envelope is observed in
dust-scattered light which originates from the central star or nebula.
The simplest situation is optically thin, isotropic scattering which
may be a reasonable approximation for some halos of proto-PNe and PNe
where local absorption effects are not dominant.  Other cases, in
which the absorption of the radiation from the centre plays an
important role, need to be considered on a case-by-case basis. The
principles, however, are the same as for those already discussed.

For the optically thin scattering case, it may be shown that $C_{\rm S}$,
the intensity ratio on the projected major and minor axes, is related
to the shape parameter $K_n$ for an envelope seen edge-on
($i=90\degr$) by:
\begin{equation}
C_{\rm S}(90) = \ln K_n \, \frac{K_n\, (1+(\ln K_n/\pi)^2)}{ K_n -(1+2(\ln
K_n/\pi)^2)}.
\end{equation}
This relation is shown as the (short) dashed line in Fig.~4. It lies
between the column density and emission measure curves. The variation
of $C_{\rm S}$ with inclination angle has been evaluated numerically, and
two examples are shown in Fig.~5. It can be seen that they are similar
in form to the curves for $C_{\rm N}$.

For both the scattering and the emission measure cases, figures
equivalent to Fig.~6 can be generated by replacing the contour values
of $C_{\rm N}$ with the equivalent values of $C_{\rm E}$ and $C_{\rm S}$ found in
Fig.~4, or directly by combining equations (6) and (7) with equation
(4). 

In the cases where dust plays a role in the observations, we
have implicitly assumed that the dust/gas ratio is constant throughout
the envelope. In fact, there is expected to be a latitude variation in
the dust-gas drift velocity because of the variation in density, and
this will produce differences between the dust and gas
envelopes. There are currently only limited observations bearing on
the question of dust-gas coupling in AGB envelopes
\citep[e.g.,][]{mau00}, but it will need to be taken into account in
detailed studies. It should not affect that main conclusions of the
present paper. 

\section{Discussion}

\subsection{Probing the presence of companions}

The prescription given above for the large scale shaping of AGB
envelopes by binary companions is potentially useful in a variety of
situations because the envelopes can be observed with different
techniques at optical, infrared, and millimetre wavelengths; they can
also be traced in the halos of proto-PNe and PNe.

In principle, the envelopes of AGB stars with known companions could be used
to check the simulations, but in practice there are currently no useful
observations available for this purpose. In the case of Mira, which
first comes to mind, the envelope appears to be completely disturbed,
from the outside by high velocity interaction with the
interstellar medium \citep{mar07}, and from the inside by the action
of bipolar jets \citep{jos00,mea09}. For other known binaries, there are no
suitable observations of the extended, undisturbed, envelopes.  We
therefore proceed on the basis that the simulations are correct,
bearing in mind that the prescription can be updated when additional
information becomes available.

One type of application is the analysis of survey observations of the
shapes of AGB envelopes in which the orientation of any binary
contribution is unknown. In this case, population syntheses based on
random orientations and assumed distributions of companion masses and
separations will be needed to evaluate the effects on the observed
shapes.  Thus the inclination can be accounted for
statistically. Since the median inclination for random orientations is
60\degr, Fig.~5 shows that the shaping signal is well preserved for
part of the population under these conditions.
 
A second type of application is the analysis of individual systems in
which the orientation of a possible underlying binary system is
already suggested by observations of other structures such as
bipolar flows. In this case the inclination can be directly taken into
account, e.g., using the curves of Fig.~5. Examples of this approach
are given in the sections that follow.

It is interesting to note that the gravitational focusing expressed by
equation (4) will respond to changes in the parameters during
evolution on the AGB. For example, an increase in the wind velocity
and an increase in separation due to mass loss leads to a secular
decrease in the degree of flattening of the envelope.  This effect
offers diagnostic possibilities, but is currently beyond observational
capabilities. The sudden transition from quasi-spherical to
axi-symmetry during PN formation (see Section 4.3), is attributable to
other physical effects.

\begin{figure}
\includegraphics[width=80mm]{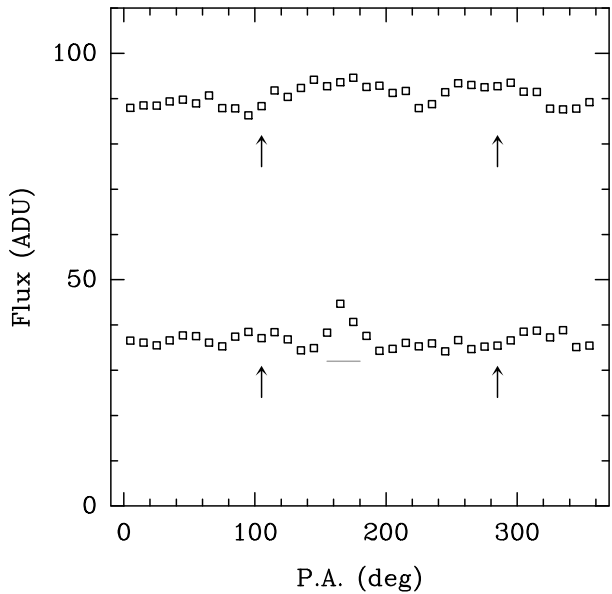}
\caption{The envelope of IRC+10216: intensity vs.\ position angle for
annuli of radii 10--20\arcsec\ (top), and 40--50\arcsec\ (bottom).  The
horizontal line near 165\degr\ marks a region in the data affected by
the halo of a bright field star. The arrows indicate P.A.s orthogonal
to the bipolar axis; no systematic equatorial enhancement is seen.
 }
\end{figure}

\subsection{IRC+10216}
As an illustration of our approach we consider the envelope of
IRC+10216, the nearest AGB star with a high mass loss rate. The
envelope has been extensively mapped in the millimetre lines of CO and
other molecular species \citep[e.g.,][]{hug88,fon06}, and it has been
imaged at high resolution in dust scattered galactic light
\citep{mau99,mau00,lea06}. These different observations all show that
the extended envelope (10--200\arcsec) is roughly circularly
symmetric.

In the centre of the envelope on arc-second size scales, HST
observations reveal a bipolar structure which is thought to represent
the early development of axi-symmetry which characterises the transition
to the proto-PNe phase \citep{ski98,mau00}.  If a companion is the cause of
the bipolarity, it must have been present during ejection of the
entire circumstellar envelope. The shape of the envelope can therefore
be used to probe the presence of the companion. 

\subsubsection{IRC+10216: envelope shape}

We quantify the large scale shape of the envelope using observations
of dust-scattered, ambient Galactic light.  Away from the centre, the
scattering optical depth is small, and the imaging provides an
excellent probe of the projected envelope shape. The observations we
use were made in the V band with the VLT (FORS1 program
64.L--0204, PI: de Laverny). The data were previously
described by \cite{lea06}, (see also \citealt{hug02}), and the
processed image is seen in their Figs.~3 and 4.  The scattered light
images of IRC+10216 show substructure in the form of multiple arcs
which have previously been discussed in detail \citep{mau00}. We are
interested here in the large scale geometry averaged over the
substructure.

After stars and galaxies were filtered from the image, the envelope
was divided into a series of annuli, 10\arcsec\ in width and centred
on the star. The annuli were further divided into 10\degr\ bins in position
angle. The pixels in each bin were then averaged to determine the
envelope intensity above the background. The results are shown in
Fig.~7 for two annuli which were the least affected by the halos of
bright stars in the field. The formal errors in the intensity are
comparable to the symbol size. The radial dimensions of the annuli are
10--20\arcsec\ and 40--50\arcsec; for a distance of 120~pc
\citep{lou93} and an expansion velocity of 14.1~km~s$^{-1}$
\citep{hug86}, they sample the envelope geometry on time scales of
400--2000~yr.

\subsubsection{IRC+10216: limits on a companion}

The bipolar structure at the centre of IRC+10216 lies at a projected
position angle of $\sim 15\degr$, and from modelling by \citet{ski98}
the inclination to the line of sight is $70\degr \pm 10\degr$.  The
expansion time scale for this feature is $\sim50$~yr, so it is very
recent \citep{mau06}. If it is caused by interaction with a companion,
the companion must have been present during the ejection of the more
extended envelope sampled in Fig.~7.

The inclination angle of the bipolar axis means that the system is
seen close to edge-on. There is, however, no evidence for enhanced
emission on the projected equator (indicated by the arrows in the
Fig.~7).  Although some of the arc fine-structure deviates from
circularity, the average distribution of matter seen in the figure is 
relatively constant with position angle. The intensity ratio
between the equator and the bipolar axis is close to 1. Even after
allowing for the inclination, $C_{\rm N} \la 1.1$.

Reference to Fig.~6 shows that these results place stringent
constraints on any hypothesised companion to IRC+10216. For example,
if the bi-polarity is caused by the onset of interaction with a
companion initially within $\sim 10$~AU of the primary, the
possibility that it can be a stellar object of mass $\ga
0.1$~M$_{\sun}$ appears to be ruled out, otherwise the envelope would
exhibit observable large scale asymmetry.  This result is even
stronger when one considers a lower value of $V_{\rm s}$ appropriate
for small separations, rather than the nominal 10~km~s$^{-1}$ used for
Fig.~6.

Because of the low limit of $C_{\rm N}$, Fig.~6 also shows that even at
distances larger than 10~AU from the primary, the mass of any
companion is still stringently constrained.

\subsection{Binary interaction in the PN transition? }

A similar situation is seen in PNe and proto-PNe where the
axi-symmetric structure of the nebulae is often surrounded by an
extended halo, which is identified with earlier mass loss of the star
on the AGB.  Numerous examples are seen in the survey observations of
\citet{cor03} and \citet{sah07}.   On the largest size scales the
dominant shaping effect of halo gas is interaction with the
interstellar medium \citep[e.g.,][]{war07}.  Closer to the star the
halo shape is expected to reflect the mass-loss geometry on the
AGB.  It can therefore place important constraints on the origins of
bipolar structure in the nebulae, and we discuss this below. It is
notable that the PNe with known close companions lack halos
\citep{bon90}, which suggests that these form before the late stages
on the AGB.

\subsubsection{Halo shapes}

There is little detailed information on the intrinsic shapes of halos, but they
must typically be close to spherically symmetric. There are few, if
any, that appear flattened, and they appear round in projection even
when the axi-symmetric structure at the centre is seen close to
edge-on. The equatorial tori, which often form part of the structure
of PNe, do not extend into the halos, and there is good evidence that
the tori develop on relatively short time scales along with the
axi-symmetry and jets \citep{hug07}.

In the case of evolved PNe, where the nebula is completely
photo-ionized by the central star, the gas pressure may play a role in
the evolution of the shape of the halo, as described in Section
3.3.2. This effect is less important on shorter time scales, and
unimportant in proto-PNe, before the onset of ionisation. Hence the
shapes of the halos of proto-PNe provide direct information on the
geometry of the mass loss before the nebula forms.

In a recent HST imaging survey of proto-PNe, \cite{sah07} found that
the majority (70\%) of the 23 objects observed have detectable halos
with discernible shapes. Essentially all of these have
axi-symmetric structures in their central regions. Most of the halos
(75\%) are observed to be round. The remainder appear elongated along
the bipolar axes of the central structure and are most probably
illuminated by (or through) the bipolar lobes; in these cases the
overall geometry of the halo is not revealed, but is consistent with
spherical symmetry and non-uniform illumination. Thus the sample as a
whole provides evidence that mass loss before the onset of
axi-symmetry, is mostly spherically symmetric.

In order to quantify the shapes of the halos we have examined the
images reported by \cite{sah07} of several proto-PNe, in which the
bipolar structure at the center has a considerable inclination angle
to the line of sight, as evidenced by their detailed shapes and
absorption lanes. These include IRAS 13357$-$6442, IRAS 17440$-$3310,
IRAS 18420$-$0512, and IRAS 19475+3119. We characterise the observed
shapes of the halos using the intensity ratio at equal distances from
the center on axes aligned with the projected equatorial and polar
axes of the central bipolar structure (avoiding diffraction spikes,
jets, etc.). These ratios, which correspond to $C_{\rm S}$ in section 3.3.3,
all lie in the range 0.6--1.2. The measured uncertainties are $\la
15\%$, and there is good agreement between the results obtained from
images with different filters. The individual inclination angles are
not known, but they are large by selection (see above). It is
reasonable to assume that they are typically $\ga 60\degr$; using the
inclination calculations of section 3.3.3 we can set an upper limit
for the edge-on value of $C_{\rm S}$ of $\la 1.5$. The corresponding column
density ratio $C_{\rm N}$ is $\sim 1.3$.  There is no evidence for any
substantially flattened halos lying in the equatorial plane of the
central bipolar structures.

The typical size scale sampled in these halo observations is $\sim
10^{17}$~cm. For a typical halo velocity of 10~km\,s$^{-1}$ the
corresponding expansion age is $\sim 3000$~yr. The ages of the central
bipolarity are typically less by factors of $\sim 5$--10 on account of
the smaller radial extents and larger expansion velocities. Such rapid
time scales are consistent with detailed studies of other proto-PNe
where the velocities and inclination angles are known
\citep[e.g.,][]{hug07}.

\subsubsection{Limits on companions}

The approximate spherical symmetry of proto-PN halos and the sudden
development of axi-symmetry in their inner regions is striking, and
can be used to constrain formation scenarios involving binary
companions.

The scenarios which have been proposed can be divided into two general
types. In the first, the formation of axi-symmetry and jets in the
AGB-PN transition is the result of relatively close interaction of the
primary with a companion, e.g., tidal spin-up by the companion, or
engulfment followed by envelope spin-up or ejection
\citep[e.g.,][]{sok94,rey99,nor06}.  This type of interaction can give
rise to an event with short time scales, consistent with observations
\citep{hug07}. In the second type of scenario, the components interact
via accretion of the AGB wind by the secondary, which produces jets
\citep[e.g.,][]{mor87,sok00}. In this case the jets may be triggered
by enhanced mass loss from the primary which produces enhanced
accretion by the secondary \citep{hug07}.

In either case, the relative sphericity of the mass loss before the
sudden PN event severely constrains the mass and separation of any
companion. For example, at separations $\sim 10$~AU, Fig.~6 shows that
for our nominal and conservative limit on asphericity of 1.3, the
effect of focusing alone places an upper limit of $\sim 0.2$~M$_\odot$
on the mass of the secondary.  This conclusion is not an artifact of
our data fitting. Models 13 and 18 (Table A1) with secondary masses
$M_{\rm s}= 0.25$~M$_\odot$ and separations 6.3 and 12.3~AU, both have
envelopes with large ($\ga3$) equator-to-pole density ratios and could
readily be distinguished as non spherically symmetric.  The upper
limit on the mass is even more stringent if one considers a more
realistic (lower) value for $V_{\rm s}$ than the nominal
10~km~s$^{-1}$ used for Fig.~6. In addition, for very close
encounters, other effects are expected to enhance or dominate mass
loss in the equatorial plane. These include the influence of tidal
effects on the mass loss of the AGB star \citep{fra01}, and the
possible formation of circumbinary disks (\citealt{van07}; see also
\citealt{fra07}).

The implications of the absence of flattened halos in proto-PNe and
PNe seem inescapable.  If relatively nearby binary companions are the
cause of the sudden development of axi-symmetry and jets during PN
formation, they must typically involve low mass stellar or sub-stellar
companions.  Only low mass objects may be present at these distances
from the primary without the tell-tale shaping of the extended
envelope; their effect remains invisible until some interaction with
the primary causes a sudden and major change in the mass loss
geometry. Even at distances as large as 20--30~AU the absence of
observable shaping of the halo constrains the companion mass $\la
0.3$--0.5~M$_\odot$, according to our nominal criterion. In fact it is
an interesting and open question whether a companion can avoid
significantly affecting the large scale shape of the AGB envelope by
gravitational focusing, and at the same time accrete enough material
to generate jets that initiate a transition to strong
axi-symmetry. This point is currently under further investigation.

These findings depend, of course, on the applicability of
the underlying simulations. Those used (MM99) are the most detailed
and comprehensive currently available. It is possible that effects not
included in the simulations such as time variable mass loss (evident
in the examples discussed above) or the dynamical role of magnetic
fields \citep[e.g,][]{vle02,hug05,sab07} may affect the shaping at a
quantitative level.  Thus updated simulations exploring such effects,
and detailed comparisons with observations of AGB stars with
companions will be important developments for the future.

\section{Conclusions}

This paper shows how the large scale geometry of AGB envelopes can be
used to probe the presence of unseen binary companions.  The presence
of a companion influences the AGB mass loss by focusing it towards the
orbital plane. The effect is frozen into the wind at the terminal
velocity, and so determines the shape of the envelope at large
distances from the star.

We have developed a prescription for the magnitude of the shaping
effect, based on the binary simulations of MM99. The prescription
provides a tool for the analysis of envelope images at optical,
infrared, and millimetre wavelengths.  In its present form it has a
number of limitations.  Most important is the limited parameter space
covered by the simulations. It also uses approximations to
characterise the intrinsic shapes (Section 3.1) and in relating these
to the binary parameters (Section 3.2). These can certainly be refined
in the future with the help of more extensive exploration of
parameter space.

Using the prescription, we have investigated the commonly observed
transition from spherically symmetric mass loss on the AGB to
axi-symmetry with jets in proto-PNe and PNe.  If this transition is
caused by interaction of the primary with a companion, the absence of
shaping effects in the extended envelope (or halo) places strong
constraints on the mass and separation of the companion. Only low mass
objects can orbit close to the primary without observable effects on
the envelope shape: they remain invisible until interaction
with the primary causes a sudden change in the geometry and dynamics
of ejection.

\section*{Acknowledgements}

We thank A.~Gruzinov for helpful discussions, and A.~Frank and the
referee,  A.~Zijlstra, for comments on the manuscript. This work was
supported in part by NSF grant AST 08-06910.

\bsp



\appendix

\section{Details of the models}

\begin{table*}
 \centering
 \begin{minipage}{140mm}
  \caption{Details of the binary and envelope parameters.}
  \begin{tabular}{@{}lrrrrrrrr@{}}
  \hline
   Model    &    $M_{\rm s}$ & $d$ & $V_{\rm s}$ & $V_{\rm t}$ &
   $\frac{n(0)}{n(50)}$  &  $\frac{n(0)}{n(80)}$  & $\theta_{\rm o}$ & $K_n$ \\
            &    (M$_{\sun}$)  & (AU) &
   \multicolumn{2}{c}{(km~s$^{-1}$)} & & & (deg) \\ 
 \hline
01 & 1.00 &  3.7 &  5  &  10  &  25.00 &  200.00 & 15.31 &356.63 \\  
03 & 1.00 &   9.5 & 11 &  16  &  3.20 &  6.60 &   42.69 &   8.23  \\
04 & 1.00 &  12.6 & 17 &  25  &  2.50 &  2.90 &   64.03 &   4.08  \\
05 & 1.00 &  25.3 & 15 &  15  &  1.30 &  1.48 &  197.20 &   1.58  \\
06 & 2.00 &  37.9 & 15 &  17  &  1.60 &  1.94 &  113.32 &   2.21  \\
07 & 1.00 &  50.5 & 15 &  15  &  1.50 &  1.70 &  136.35 &   1.93  \\
10 & 1.00 &   6.3 &  9 &  16  &  4.60 & 16.00 &   30.75 &  18.67  \\
11 & 1.00 &  12.6 & 12 &  16  &  1.60 &  2.00 &  110.81 &   2.25  \\
12 & 1.00 &  12.3 &  7 &  10  & 32.00 & 83.00 &   16.16 & 262.13  \\
13 & 0.25 &  12.3 &  7 &  10  &  2.60 &  3.20 &   59.99 &   4.48  \\
14 & 1.00 &  24.6 &  9 &  10  &  1.90 &  2.70 &   79.21 &   3.11  \\
15 & 0.50 &   3.7 &  5 &  10  &  7.50 & 80.00 &   21.28 &  68.61  \\
16 & 0.50 &   9.5 & 11 &  15  &  1.70 &  2.60 &   88.82 &   2.75  \\
17 & 0.50 &   6.3 &  9 &  15  &  2.90 &  4.00 &   52.06 &   5.63  \\
18 & 0.25 &   6.3 &  9 &  15  &  2.30 &  2.60 &   70.89 &   3.56  \\
\hline
\end{tabular}
\end{minipage}
\end{table*}

\subsection{Model parameters}

The simulations that we use to relate envelope shape to binary
parameters are taken from MM99. Details of the models are given in
Table A1. We omit their model 2 which includes the effects of spin-up,
and their models 8 and 9 which have wind velocities at the secondary
that are much larger than expected in AGB envelopes.  Table A1
includes the binary parameters, the density ratio between the orbital
plane and latitudes 50\degr\ and 80\degr, and our estimates of the
shape parameters $\theta_{\rm o}$ and $K_n$ discussed in section 3.1.

\subsection{Relation of envelope shape to binary parameters}

The fit of equation~(4) to the model data is shown in Fig.~1, together
with the one sigma confidence limits. A higher order curve does not
significantly improve the fit. There is a clear outlier (model 12,
Table A1), so we characterise the scatter by the median absolute
deviation, which is 0.19 dex. This translates into a scatter in
observed quantities such as the column density contrast $C_{\rm N}-1$
(viewed edge-on) of $\sim 25$\%. There are several factors that may
contribute to the scatter of the data points in Fig.~1.  (1)
Limitations in the model calculations, e.g., the influence of finite
simulation volumes, and different degrees to which the models reach
exact steady state in the far field. (2) Variation of subordinate
parameters, including the terminal velocity and details of the wind
acceleration curve. (3)  Additional physical effects such as residual
spiral structure and resonance
between the period and the wind crossing time of the orbit.  (4)
Deviations from the power law of equation (4), which may vary over
the parameter space.

The simulation grid is too sparse to explore the parameter space in
detail. However, it is possible to check whether the data are consistent
with the combination of parameters used in equation (4), by replacing
the independent variable with the more general form of $\log M_{\rm s}+
\gamma\log V_{\rm s}+\delta \log d$.  If we fit the whole data set we
obtain the solution $\gamma=-2.41 \pm 0.90$ and $\delta =
-0.50\pm0.30$, and if we omit the outlier we obtain $\gamma= -2.11 \pm
0.68$ and $\delta = -0.81 \pm 0.29$. The latter solution gives
significantly smaller residuals and is preferred, but both solutions
are consistent with the combination of parameters used in equation (3)
and support the analysis of Section 3.2. A more extensive grid of
models would be useful to explore the parameter space in detail.

\subsection{Comparisons}

To our knowledge there are no independent simulations that can be
directly compared with the detailed results of MM99. The models by
\cite{gaw02} show roughly similar levels of gravitational focusing,
but they involve some important physical differences concerning the
geometry of the mass loss and the role of cooling. Unlike the models
of MM99, most of those of \cite{gaw02} include initially anisotropic
mass loss. The one exception is their model G, which has initially
isotropic mass loss, with $M_{\rm p}=1.0$~M$_{\sun}$, $M_{\rm s}= 0.6$~M$_{\sun}$,
$V_{\rm s}= 9$~km~s$^{-1}$, and $d=9.6$~AU. From the latitude-density
profile of this model, we find that $C_{\rm N}(90)= 1.4$.  This, however, is
expected to be an underestimate of the density contrast. \citet{gaw02}
point out that the absence of radiation cooling in all their models
leads to an underestimate of the density enhancement in the equatorial
plane because the inflow is partly suppressed by high pressure gas
that forms around the secondary.  Consistent with this expectation, if
we use the binary parameters of model G in equations (4) and (5), we
predict a higher contrast ($C_{\rm N}=2.4$) as expected.  More detailed and
extensive cross-checks on the model results would clearly be useful.

\subsection{Computational domain}
The radial extent of the computational domain of the MM99 simulations
ranges from $\sim$ $1\times10^{15}$ to $3\times10^{16}$~cm. The radii
of the observations of IRC+10216 shown in Fig.~7 are near the upper
end of this range, and those of the proto-PNe halos are up to an order
of magnitude larger. For envelopes in steady state at the terminal
velocity in each direction, the density profile remains constant with
increasing radius, as assumed. The gas at large distances from the
star is cool and the turbulent velocity (measured in few cases, e.g.,
\citealt{hug86}) is low, so that these should not have significant
dispersal effects. To the extent that the simulations have not reached
exact steady state at the terminal velocity within the computational
domain, there may be residual changes of shape farther out, but these
are expected to be small. Simulations on extended domains are needed
to quantify this.

\label{lastpage}


\begin{thebibliography}{}


\bibitem[\protect\citeauthoryear{Bond \& Livio}{1990}]{bon90} 
Bond H.~E., Livio M., 1990, ApJ, 355, 568 

\bibitem[\protect\citeauthoryear{Chiu et al.}{2006}]{chi06}
Chiu P.-J., Hoang C.-T., Dinh-V-Trung, Lim J., Kwok S., Hirano
N., Muthu, C., 2006, ApJ, 645, 605


\bibitem[\protect\citeauthoryear{Corradi et al.}{2003}]{cor03}
Corradi R.~L.~M., Sch{\"o}nberner D., Steffen M., Perinotto, M.,
2003, MNRAS, 340, 417

\bibitem[\protect\citeauthoryear{de Marco}{2006}]{dem06} 
de Marco O., 2006, in Barlow M. J., Mendez R. H., eds, Planetary 
Nebulae in our Galaxy and Beyond, IAU Symp.\ 234, CUP, Cambridge, p.~111 

\bibitem[\protect\citeauthoryear{Fong et al.}{2006}]{fon06} 
Fong D., Meixner M., Sutton E.~C., Zalucha A., Welch W.~J.,
2006, ApJ, 652, 1626

\bibitem[\protect\citeauthoryear{Frankowski \& Tylenda}{2001}]{fra01}
Frankowski A., Tylenda R., 2001, A\&A, 367, 513 

\bibitem[\protect\citeauthoryear{Frankowski \& Jorissen}{2007}]{fra07} 
Frankowski A., Jorissen A.,2007, Baltic Astronomy, 16, 104 

\bibitem[\protect\citeauthoryear{Gawryszczak et al.}{2002}]{gaw02}
Gawryszczak A. J., Mikolajewska K., R\'{o}zyczka M., 2002, A\&A,
385, 205

\bibitem[\protect\citeauthoryear{Han et al.}{1995}]{han95}
Han Z., Podsiadlowski P., Eggleton P.~P., 1995, MNRAS, 272, 800 

\bibitem[\protect\citeauthoryear{Huggins}{2007}]{hug07} 
Huggins P.~J., 2007, ApJ, 663, 342 

\bibitem[\protect\citeauthoryear{Huggins \& Healy}{1986}]{hug86}
Huggins P.~J., Healy A.~P., 1986, ApJ, 304, 418 

\bibitem[\protect\citeauthoryear{Huggins \& Manley}{2005}]{hug05} 
Huggins P.~J., Manley S.~P., 2005, PASP, 117, 665

\bibitem[\protect\citeauthoryear{Huggins \& Mauron}{2002}]{hug02} 
Huggins P.~J., Mauron N., 2002, A\&A, 393, 273 

\bibitem[\protect\citeauthoryear{Huggins et al.}{1988}]{hug88}
Huggins P.~J., Olofsson H., Johansson L.~E.~B., 1988, ApJ, 332,
1009

\bibitem[\protect\citeauthoryear{Josselin et al.}{2000}]{jos00}
Josselin E., Mauron N., Planesas P., Bachiller R., 2000, A\&A, 362, 255 

\bibitem[\protect\citeauthoryear{Jorissen}{2003}]{jor03} Jorissen, A.\
2003, in Habing H. J., Olofsson H., eds, Asymptotic giant branch stars,
Springer, Berlin, p. 461

\bibitem[\protect\citeauthoryear{Le{\~a}o et al.}{2006}]{lea06} 
Le{\~a}o I.~C., de Laverny P., M{\'e}karnia D., de Medeiros J.~R.,
Vandame B., 2006, A\&A, 455, 187 

\bibitem[\protect\citeauthoryear{Loup et al.}{1993}]{lou93} 
Loup C., Forveille T., Omont A., Paul J.~F., 1993, A\&A Supp., 99, 291 

\bibitem[\protect\citeauthoryear{Martin et al.}{2007}]{mar07} 
Martin D.~C., et al., 2007, Nature, 448, 780

\bibitem[\protect\citeauthoryear{Mastrodemos}{1998}]{mas98} 
Mastrodemos N.\ 1998, Ph.D.~Thesis,  

\bibitem[\protect\citeauthoryear{Mastrodemos \& Morris}{1999}]{mas99}
Mastrodemos N., Morris M., 1999, ApJ, 523, 357 (MM99)

\bibitem[\protect\citeauthoryear{Mauron \& Huggins}{1999}]{mau99}
Mauron N., Huggins P. J., 1999, A\&A 349, 203

\bibitem[\protect\citeauthoryear{Mauron \& Huggins}{2000}]{mau00} 
Mauron N., Huggins P.~J., 2000, A\&A, 359, 707 

\bibitem[\protect\citeauthoryear{Mauron \& Huggins}{2006}]{mau06}
Mauron N., Huggins P.~J., 2006, A\&A, 452, 257

\bibitem[\protect\citeauthoryear{Meaburn et al.}{2009}]{mea09} Meaburn
J., Lopez J.~A., Boumis P., Lloyd M., Redman, M.~P., 2009,
arXiv:0903.1966

\bibitem[\protect\citeauthoryear{Mellema \& Frank}{1995}]{mel95} 
Mellema  G., Frank A., 1995, MNRAS, 273, 401 


\bibitem[\protect\citeauthoryear{Miszalski et al.}{2009}]{mis09} 
Miszalski B., Acker A., Moffat A.~F.~J., Parker Q.~A., Udalski A.,
2009, arXiv:0901.4419

\bibitem[\protect\citeauthoryear{Morris}{1987}]{mor87}
Morris M., 1987, PASP, 99, 1115

\bibitem[\protect\citeauthoryear{Morris et al.}{2006}]{mor06} 
Morris M., Sahai R., Matthews K., Cheng J., Lu J., Claussen M., \&
S{\'a}nchez-Contreras C., 2006, in Barlow M.~J., Mendez R.~H., eds,
Planetary Nebulae in our Galaxy and Beyond, IAU Symp.\ 234, CUP,
Cambridge, p.~469

\bibitem[\protect\citeauthoryear{Nordhaus \& Blackman}{2006}]{nor06}
Nordhaus J., Blackman E.~G., 2006, MNRAS, 370, 2004


\bibitem[\protect\citeauthoryear{Reyes-Ruiz \& L{\'o}pez}{1999}]{rey99}
Reyes-Ruiz M., L{\'o}pez J.~A., 1999, ApJ, 524, 952

\bibitem[\protect\citeauthoryear{Sabin et al.}{2007}]{sab07} 
Sabin L., Zijlstra A.~A., Greaves J.~S., 2007, MNRAS, 376, 378 

\bibitem[\protect\citeauthoryear{Sacuto et al.}{2008}]{sac08}
Sacuto S., Jorissen A., Cruzal{\`e}bes P., Chesneau O., Ohnaka
K., Quirrenbach A., Lopez B., 2008, A\&A, 482, 561

\bibitem[\protect\citeauthoryear{Sahai et al.}{2007}]{sah07} 
Sahai R., Morris M., 
S{\'a}nchez Contreras C., Claussen M. 2007, AJ, 134, 2200 


\bibitem[\protect\citeauthoryear{Skinner et al.}{1998}]{ski98} 
Skinner C.~J., 
Meixner M., Bobrowsky M., 1998, MNRAS, 300, L29 

\bibitem[\protect\citeauthoryear{Soker \& Livio}{1994}]{sok94}
Soker N.,  Livio, M., 1994, ApJ, 421, 219

\bibitem[\protect\citeauthoryear{Soker \& Rappaport}{2000}]{sok00}
Soker N., Rappaport S., 2000, ApJ, 538, 241

\bibitem[\protect\citeauthoryear{Theuns \& Jorissen}{1993}]{the93}
Theuns T., Jorissen A., 1993, MNRAS, 265, 946

\bibitem[\protect\citeauthoryear{Vlemmings et al.}{2002}]{vle02}
Vlemmings W.~H.~T., Diamond P.~J., van Langevelde H.~J., 2002,
A\&A, 394, 589  

\bibitem[\protect\citeauthoryear{Van Winckel}{2007}]{van07} 
Van Winckel H., 2007, Baltic Astronomy, 16, 112 

\bibitem[\protect\citeauthoryear{Wareing et al.}{2007}]{war07} 
Wareing C.~J., Zijlstra A.~A., O'Brien, T.~J., 2007, MNRAS, 382, 1233 

\end{thebibliography}
\end{document}